\documentclass[conference]{IEEEtran_NGOPERA26}
\IEEEoverridecommandlockouts

\usepackage{cite}
\usepackage{amsmath,amssymb,amsfonts}
\usepackage{algorithmic}
\usepackage{graphicx}
\usepackage{textcomp}
\def\BibTeX{{\rm B\kern-.05em{\sc i\kern-.025em b}\kern-.08em
    T\kern-.1667em\lower.7ex\hbox{E}\kern-.125emX}}
\usepackage{algorithm}
\usepackage{array}
\usepackage[caption=false,font=normalsize,labelfont=sf,textfont=sf]{subfig}
\usepackage{tabularx}
\usepackage{stfloats}
\usepackage{url}
\usepackage{verbatim}
\usepackage{booktabs,multirow}
\usepackage{array,siunitx}
\usepackage{doi}
\newcolumntype{A}{>{\raggedright\arraybackslash}p{0.55\linewidth}}
\newcolumntype{L}{>{\raggedright\arraybackslash}X}
\newcolumntype{C}{>{\centering\arraybackslash}X}
\newcolumntype{D}[1]{>{\centering\arraybackslash}p{#1}}
\newcolumntype{V}{>{\raggedleft\arraybackslash}p{1.6cm}}
\usepackage{acro}[=v2]
\usepackage[table]{xcolor}
\DeclareAcronym{E2SM-RC}{
    short=E2SM-RC,
    long=E2 Service Model-RAN Control
}

\DeclareAcronym{ORAN}{
    short=O-RAN,
    long=Open Radio Access Network,
    short-plural-form=O-RAN
}

\DeclareAcronym{RIC}{
    short=RIC,
    long=RAN Intelligent Controller
}

\DeclareAcronym{Near-RT}{
    short=Near-RT,
    long=Near-Real Time
}

\DeclareAcronym{Non-RT}{
    short=Non-RT,
    long=Non-Real Time
}

\DeclareAcronym{MRO}{
    short=MRO,
    long=Mobility Robustness Optimization
}

\DeclareAcronym{EE}{
    short=EE,
    long=Energy Efficiency
}

\DeclareAcronym{MLB}{
    short=MLB,
    long=Mobility Load Balancing
}

\DeclareAcronym{OTIC}{
    short=OTIC,
    long=Open Test and Integration Center
}

\DeclareAcronym{CM}{
    short=ConMit,
    long= Conflict Mitigation
}

\DeclareAcronym{ML}{
    short=ML,
    long=Machine Learning
}

\DeclareAcronym{RL}{
    short=RL,
    long=Reinforcement Learning
}

\DeclareAcronym{CMF}{
    short=CMF,
    long=Conflict Mitigation Framework
}

\DeclareAcronym{CD}{
    short=CD,
    long=Conflict Detection
}

\DeclareAcronym{CR}{
    short=CR,
    long=Conflict Resolution
}

\DeclareAcronym{eMBB}{
    short=eMBB,
    long=enhanced Mobile Broadband
}

\DeclareAcronym{gNB}{
    short=gNB,
    long=gNodeB
}

\DeclareAcronym{BS}{
    short=BS,
    long=base station,
    short-plural-form=BSes
}

\DeclareAcronym{UE}{
    short=UE,
    long=User Equipment 
}

\DeclareAcronym{RLF}{
    short=RLF,
    long=Radio Link Failure 
}

\DeclareAcronym{CB}{
    short=CB,
    long=Call Blockage 
}

\DeclareAcronym{TTT}{
    short=TTT,
    long=time-to-trigger 
}

\DeclareAcronym{CIO}{
    short=CIO,
    long=Cell Individual Offset
}

\DeclareAcronym{O-RAN SC}{
    short=O-RAN SC,
    long=O-RAN Software Community 
}

\DeclareAcronym{RMR}{
    short=RMR,
    long=RIC Message Router 
}

\DeclareAcronym{SDL}{
    short=SDL,
    long=Shared Data Layer
}

\DeclareAcronym{E2AP}{
    short=E2AP,
    long=E2 Application Protocol
}

\DeclareAcronym{TS}{
    short=TS,
    long=Traffic Steering 
}

\DeclareAcronym{SM}{
    short=SM,
    long=Slice Management
}

\DeclareAcronym{MIMO}{
    short=MIMO,
    long=Multiple Input Multiple Output 
}

\DeclareAcronym{KPM}{
    short=KPM,
    long=Key Performance Measurement
}

\DeclareAcronym{API}{
    short=API,
    long=Application Programming Interface
}

\DeclareAcronym{srsRAN}{
    short=srsRAN,
    long=Software Radio Systems RAN 
}

\DeclareAcronym{COTS}{
    short=COTS,
    long=Commercial Off-The-Shelf 
}

\DeclareAcronym{KPI}{
    short=KPI,
    long=Key Performance Indicator
}

\DeclareAcronym{PRB}{
    short=PRB,
    long=Physical Resource Block
}

\DeclareAcronym{QoS}{
    short=QoS,
    long=Quality of Service
}

\DeclareAcronym{RAN}{
    short=RAN,
    long=Radio Access Network
}

\DeclareAcronym{RF}{
    short=RF,
    long=Radio Frequency
}

\DeclareAcronym{SDR}{
    short=SDR,
    long=Software-defined Radio
}

\DeclareAcronym{CU}{
    short=O-CU,
    long=Open Central Unit
}

\DeclareAcronym{RU}{
    short=O-RU,
    long=Open Radio Unit
}

\DeclareAcronym{DU}{
    short=O-DU,
    long=Open Distributed Unit
}

\DeclareAcronym{gNodeB}{
    short=gNB,
    long=gNodeB
}

\DeclareAcronym{SMO}{
    short=SMO,
    long=Service Management and  Orchestration
}

\DeclareAcronym{E2SM}{
    short=E2SM,
    long=E2 Service Model
}

\DeclareAcronym{UHD}{
    short=UHD,
    long=\acs{USRP} Hardware Driver
}

\DeclareAcronym{USRP}{
    short=USRP,
    long=Universal Software Radio Peripheral
}

\DeclareAcronym{VM}{
  short=VM,
  long=Virtual Machine
}

\DeclareAcronym{MNO}{
    short=MNO,
    long=Mobile Network Operator
}

\DeclareAcronym{DL}{
    short=DL,
    long=Downlink
}

\DeclareAcronym{UL}{
    short=UL,
    long=Uplink
}

\DeclareAcronym{SD}{
    short=SD,
    long=Standard Deviation
}

\DeclareAcronym{OTA}{
    short=OTA,
    long=Over The Air
}

\DeclareAcronym{AI}{
    short=AI,
    long=Artificial Intelligence 
}

\DeclareAcronym{ANN}{
    short=ANN,
    long=Artificial Neural Network 
}

\DeclareAcronym{DQN}{
    short=DQN,
    long=Deep Q-Network
}

\DeclareAcronym{DDQN}{
    short=DDQN,
    long=Double Deep Q-Network
}

\DeclareAcronym{SELU}{
    short=SELU,
    long=Scaled Exponential Linear Unit
}

\DeclareAcronym{RELU}{
    short=ReLU,
    long=Rectified Linear Unit
}

\DeclareAcronym{MLP}{
    short=MLP,
    long=Multilayer Perceptron
}

\DeclareAcronym{PPO}{
    short=PPO,
    long=Proximal Policy Optimization
}

\DeclareAcronym{KL}{
    short=KL,
    long=Kullback-Leibler
}

\DeclareAcronym{ACCORD}{
    short=ACCoRD,
    long=Actor-Critic Conflict Resolution with Deep learning
}

\DeclareAcronym{COMIX}{
    short=COMIX,
    long=generalized Conflict Management scheme for Multi-Channel Power Control in O-RAN xApps
}

\DeclareAcronym{QACM}{
    short=QACM,
    long=QoS-Aware Conflict Mitigation
}
\ExplSyntaxOn
\prop_new:N \l__acro_foreign_format_prop
\ExplSyntaxOff

\begin{document}

\title{ACCoRD: Actor-Critic Conflict Resolution\\with Deep learning for O-RAN xApps}

\author{
\IEEEauthorblockN{Cezary~Adamczyk}
\IEEEauthorblockA{\textit{Institute of Radiocommunications} \\
\textit{Poznan University of Technology}\\
Poznań, Poland \\
cezary.adamczyk@doctorate.put.poznan.pl}
\and
\IEEEauthorblockN{Adrian~Kliks}
\IEEEauthorblockA{\textit{Institute of Radiocommunications} \\
\textit{Poznan University of Technology }\\
Poznań, Poland}
\IEEEauthorblockA{\textit{Lule\aa\ University of Technology} \\
Lule\aa, Sweden  \\
adrian.kliks@put.poznan.pl}

\thanks{This is the author’s version of an article that has been published in the proceedings of the 2026 IEEE INFOCOM conference.}}

\IEEEpubid{Copyright © 2026 IEEE. Personal use is permitted, but republication/redistribution requires IEEE permission.}

\maketitle

\begin{abstract}
\ac{CM} is a crucial part of intelligent network control in \acp{ORAN}. In this paper, we propose a method named \acs{ACCORD} to resolve detected control conflicts in Near-Real Time RAN Intelligent Controller using a \ac{CR} Agent with an \ac{ANN} trained with a reinforcement learning algorithm \acs{PPO}-Clip. The implemented \ac{ANN} analyzes data about the network and conflicting control decisions to infer optimal \ac{CR} actions. The \ac{CR} Agent gathers feedback from the network after each resolved conflict to assess its efficiency and adjust the \ac{ANN}'s weights during batch training. The evaluation of the proposed approach is based on simulation data. A new methodology for evaluating \ac{CR} solutions is proposed. Results show that the proposed \ac{ANN}-based method improves on the efficiency of rule-based approaches by significantly reducing negative network events caused by conflicting control decisions in medium and high traffic scenarios.
\end{abstract}

\begin{IEEEkeywords}
conflict resolution, reinforcement learning, conflict mitigation, Near-RT RIC, O-RAN, xApp.
\end{IEEEkeywords}

\acresetall

\section{Introduction}
\label{sec:intro}
\IEEEPARstart{O}{ne} of the promises of Open Radio Access Networks \mbox{(O-RAN)} is to enable multi-vendor \acs{RAN} with robust network optimization solutions \cite{o-ran-understanding-2022}.
Mobile Network Operators (MNOs) that deploy \acs{ORAN}-compliant networks can mix and match hardware and software components from numerous vendors.
Such heterogeneity of the RAN, if not managed properly, can lead to harmful interoperability issues.
As standards describing \ac{ORAN} mature over time and complexity of \ac{ORAN} deployments advances, \ac{CM} in \ac{ORAN}'s RAN Intelligent Controllers (RICs) becomes an increasingly important research topic.

The \ac{ORAN} architecture is designed with native support for \ac{AI} and \ac{ML} techniques.
These methods are ideally suited for \ac{CM}, where the execution of complex processing logic against large datasets can address challenges such as ensuring the reliability of conflict detection mechanisms and determining optimal \ac{CR} logic.
The complexity and dynamic nature of \ac{ORAN} make efficient conflict mitigation using static, rule-based conventional approaches challenging.
These \ac{AI}/\ac{ML} techniques are envisioned as enablers for robust \ac{CM} measures, which can adapt to ever-changing network conditions and the evolving suite of deployed applications.

\ac{AI}/\ac{ML} enables the switch from conflict resolution using static rules (e.g., fixed application priorities) to dynamic optimization.
The expected approach involves an \ac{AI} model acting as the core of the resolution logic.
This model ingests a wide range of parameters describing network state and the detected conflict to decide on the optimal set of resolution actions.
Such a system can benefit from \ac{ML} approaches, like \ac{RL}, allowing the \ac{AI} model to adapt its operation and improve its decision-making over time based on the reward signal from the \ac{ORAN} environment.

\IEEEpubidadjcol

\subsection{Related works}
Although it is a relatively new field of research, there is already a variety of work on \ac{CM} in \ac{ORAN}.
One of the first steps towards achieving a standardized \ac{CM} approach was the introduction of the \ac{CMF} by Adamczyk and Kliks \cite{oa-caak-cmf-2023}, which defined architectural extensions for the \ac{Near-RT} \ac{RIC} but relied on simple prioritization methods.
Other works, such as \ac{QACM} \cite{cm-qacm-2024}, propose rule-based methods focusing on QoS targets but require non-standard inputs from xApps, making them impractical for multi-vendor deployments.
Similarly, digital twin-based approaches like \ac{COMIX} \cite{cm-comix-2025} face challenges regarding computational delays and model accuracy.
Alternative approaches like PACIFISTA \cite{cm-pacifista-2025} operate from the \ac{SMO} level, resolving conflicts by disabling applications entirely, which removes both negative and positive impacts.
Attempts at \ac{ML}-powered solutions, such as xApp distillation \cite{cm-distillation-2024}, require complex operational efforts to compile multiple policies into a single agent, lacking "plug \& play" capability.

\subsection{Problem statement and proposed solution}
\label{sec:problem}
To address the identified gap in robust \ac{CR} methods for \ac{CM} in the \ac{Near-RT} \ac{RIC}, we propose \ac{ACCORD} (\acl{ACCORD}), a novel \ac{CR} solution based on \ac{ANN}, designed for deployment within the \ac{CMF}.
\ac{ACCORD} functions as the main decision-making entity within the \ac{CR} Agent.
By integrating an \ac{ANN}, \ac{ACCORD} analyzes control conflicts by processing data regarding conflicting decisions and current network decisions to infer target resolution actions.
While conflicts in O-RAN can be categorized into direct, indirect, and implicit types \cite{o-ran-wg3-Near-RT-RIC-con-mit}, \ac{ACCORD} is evaluated against indirect control conflicts in this study. However, its feature encoding design allows it to scale to other conflict types in the future.

\ac{ACCORD} operates by intercepting conflict reports from the \ac{CD} Agent, collecting the necessary context information, and pre-processing inputs for its \ac{ANN}.
For each detected conflict, the \ac{ANN} infers a set of resolution actions, which are applied to the conflicting control decisions.
An actor-critic architecture trained via the \ac{PPO}-Clip algorithm is employed to enable \ac{ANN}'s adaptation to dynamic network conditions.

The operation of \ac{ACCORD} follows a specific data flow:
\begin{enumerate}
    \item \textbf{Control decision issuance:} xApps submit their control decisions as E2 Control messages.
    \item \textbf{Conflict detection:} The \ac{CD} Agent analyzes the messages. If a conflict is detected, a conflict report is generated and sent to the \ac{CR} Agent.
    \item \textbf{Computation of policy logits:} \ac{ACCORD} preprocesses conflict data with global features and feeds it through the \ac{ANN} to compute \ac{CR} action logits and a critic value.
    \item \textbf{Execution of conflict resolution actions:} \ac{ACCORD} interprets the logits to determine resolution actions (e.g., modify, reject), applies them, and submits the post-resolution decisions.
    \item \textbf{Post-resolution observation:} After resolving a conflict, \ac{ACCORD} triggers a post-resolution observation of the network to calculate the reward for \ac{RL} training.
\end{enumerate}

\begin{figure*}[!t]
\centerline{\includegraphics[width=0.65\textwidth]{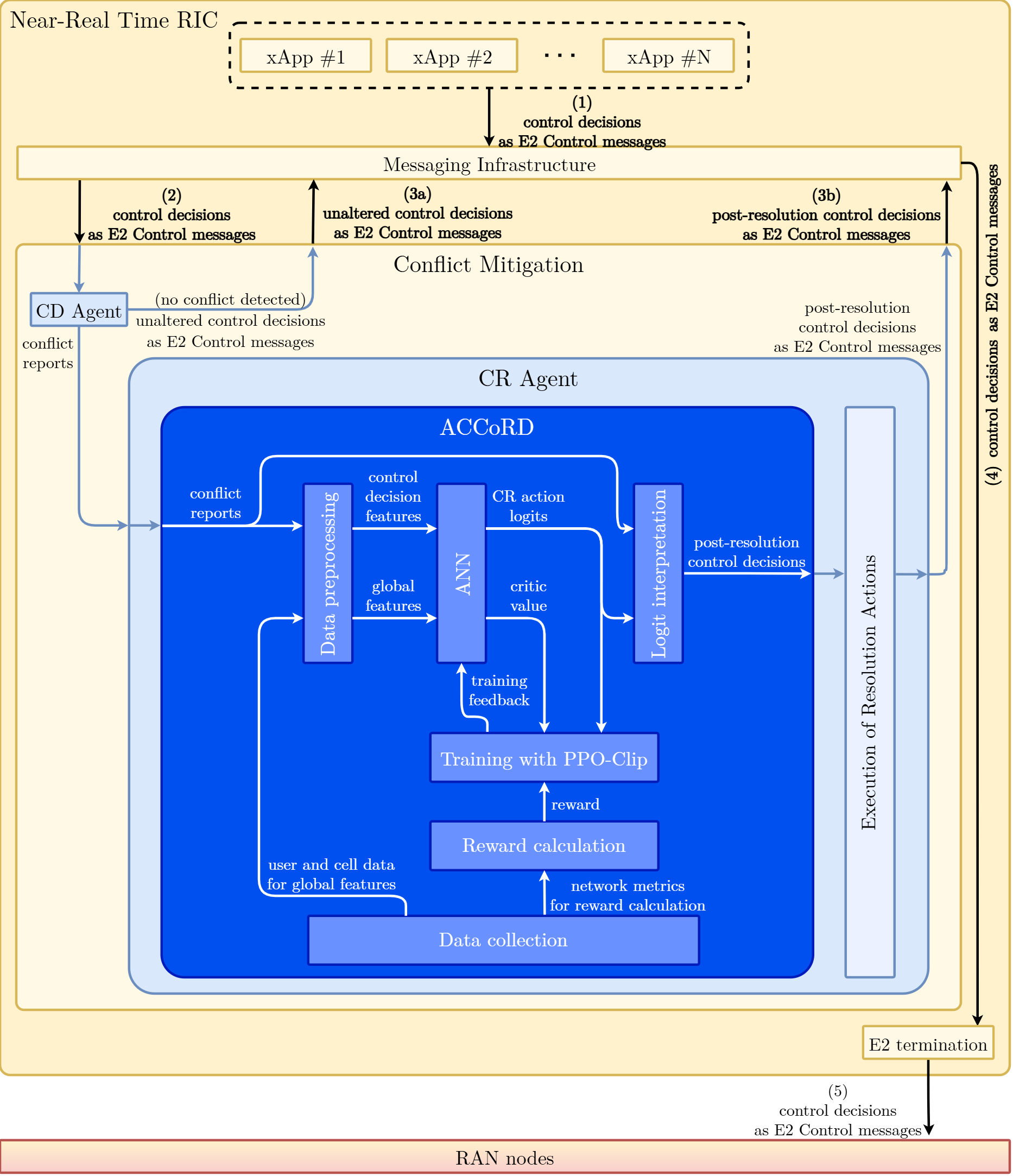}}
\caption{Detailed data flow for ACCoRD operation}
\label{fig:accord-dataflow}
\end{figure*}

To visualize these interactions, the detailed data flow for \ac{ACCORD} is presented in Figure~\ref{fig:accord-dataflow}.

\section{Design of the ANN in ACCoRD}
\label{sec:ANN}
Using \ac{AI}/\ac{ML} techniques, such as an \ac{ANN} trained using one of the \ac{RL} algorithms, \ac{CM} can be performed without explicitly defining a specific resolution logic.
The efficiency of an \ac{AI}/\ac{ML}-based solution relies heavily on its design: input feature selection, \ac{ANN} architecture, utilized \ac{ML} algorithm, and the reward function.

\subsection{ANN inputs}
The input architecture for \ac{ACCORD}'s \ac{ANN} is designed to simultaneously process the aggregate environment state and the details of conflicting control decisions.
The \ac{ANN} accepts inputs divided into two primary categories: global features, which provide contextual information regarding the target cell and its active users, and control decision features, which encode the specific control decisions involved in the conflict.

\subsubsection{Control decision features}
Each inference step processes data describing up to $N_{\mathrm{ConfDec}}$ conflicting control decisions.
The \ac{ANN} ingests these as a fixed-size, slot-based input tensor of dimensions $(N_{\mathrm{ConfDec}}, 5)$.
Each slot $i$ corresponds to a decision head and contains a feature vector $\mathbf{x}_i = [\mathbf{e}_{type} \parallel v'_{param}]$.
\begin{itemize}
    \item \textbf{Parameter type embedding ($\mathbf{e}_{type}$):} a 4-dimensional one-hot vector identifying the parameter type (none, CIO, TTT, hysteresis). This ensures the \ac{ANN} interprets parameters based on learned categorical relationships rather than arbitrary numeric representations.
    \item \textbf{Parameter value ($v'_{param}$):} the numeric value of the control parameter, normalized to the range $[0, 1]$ using min-max scaling specific to the parameter type bounds.
\end{itemize}
If a parameter type is absent from a conflict, its slot is filled with the \texttt{none} embedding and a zero value, and masked during processing.

\subsubsection{Global features}
Global features anchor the decision-making in the current network context. This vector consists of 11 floating-point values, explicitly normalized to ensure training stability:
\begin{itemize}
    \item \textbf{Cell state:} includes current values for hysteresis, CIO, and TTT (normalized by their maximum configurable values), and KPIs for availability (load $<100\%$), handover stability (non-ping-pong ratio), and total throughput (normalized by a constant of $100$ Mbps).
    \item \textbf{User state:} aggregated state of users connected to the target cell, including movement speed (normalized by $100$ m/s), requested bitrate (normalized by $1$ Gbps), satisfaction ratio, connection success rate, and throughput (normalized by $1$ Mbps).
\end{itemize}

\subsection{ANN outputs}
The output layer defines the action space and provides auxiliary outputs for training.
\ac{ACCORD} generates three sets of outputs: \ac{CR} action logits, a validity mask, and a critic value.

\subsubsection{CR action logits}
The \ac{ANN} utilizes a multi-head output structure with $N_{\mathrm{ConfDec}}$ decision heads.
Each head outputs a vector of $N_A=4$ logits, representing the preference for specific resolution actions. These logits are transformed via Softmax into a probability distribution $P(a_i)$.
The discrete action space $\mathcal{A}$ consists of:
\begin{itemize}
    \item \texttt{NO\_MODIFICATION}: The decision is applied as is.
    \item \texttt{REJECTION\_WITH\_COOLDOWN}: The decision is rejected, and the source xApp is put on a cooldown.
    \item \texttt{INCREASE\_1}/\texttt{DECREASE\_1}: The target parameter is incremented/decremented by a single logical step.
\end{itemize}
The \texttt{INCREASE\_1} and \texttt{DECREASE\_1} actions map to the discrete values supported by the specific xApp (e.g., specific TTT values defined in 3GPP standards).
For parameters with non-linear spacing, such as TTT, this method ensures that \ac{ACCORD} strictly adheres to valid configuration steps without requiring separate logic for each parameter type.

\subsubsection{Validity mask}
Because the \ac{ANN} utilizes a fixed number of decision heads ($N_{\mathrm{ConfDec}}$) to process a variable number of conflicting decisions, not all output heads produce valid data for every inference step.
To address this, the \ac{ANN} outputs a validity mask—a boolean tensor of dimension $N_{\mathrm{ConfDec}}$.
This mask corresponds to the input slots and identifies which decision heads represent real conflicting control decisions and which correspond to padding.
During post-processing, this mask is applied to the output logits; invalid slots are masked with negative infinity ($-\infty$).
This mechanism ensures that their resulting probability after the Softmax operation is effectively zero.
Consequently, the \ac{ANN}'s policy is derived from valid slots only, preventing training on padding noise.

\subsubsection{Critic value}
The critic head outputs a scalar state-value estimate $V(s)$, representing the expected cumulative reward, used to compute advantages during training.
Mathematically, the critic approximates the expected return starting from state $s$ following policy $\pi_\theta$:
\begin{equation}
    V(s) = \mathbb{E}\left[ \sum_{t=0}^{\infty} \gamma^t r_t \mid s_0 = s, \pi_\theta \right]
    \label{eq:critic-value}
\end{equation}
where $\gamma$ is the discount factor and $r_t$ is the reward at time $t$.

\subsection{ANN structure}
\label{subsec:ANN-structure}
The \ac{ANN} implemented in \ac{ACCORD} is a feed-forward actor-critic network designed to handle a variable number of conflicting decisions while maintaining context awareness.
Its internal architecture comprises three main functional blocks:

\subsubsection{Control decision feature encoder}
This component is a shared two-layer \ac{MLP} with \ac{RELU} activation.
It processes the normalized input feature ($v'_{param}$) for each conflicting decision independently.
This encoding transforms the scalar input into a higher-dimensional latent representation, which allows the subsequent layers to process abstract features rather than raw values.

\subsubsection{Control decision heads (actor)}
The network contains ${N_{\mathrm{ConfDec}}}$ distinct control decision heads.
To ensure that the resolution of one conflict considers the context of others, each head receives a composite input vector consisting of:
\begin{itemize}
    \item The encoded features of the specific control decision.
    \item The parameter type embedding ($\mathbf{e}_{type}$).
    \item The vector of global features (cell and user state).
    \item A mean-pooled summary of the encoded features from all \textit{other} conflicting decisions.
\end{itemize}
This context-aware design allows the head to output a conflict resolution action that is locally optimal for the specific decision but globally consistent with the overall conflict scenario.

\subsubsection{Critic}
\begin{figure*}[!htb]
\centerline{\includegraphics[width=0.8\textwidth]{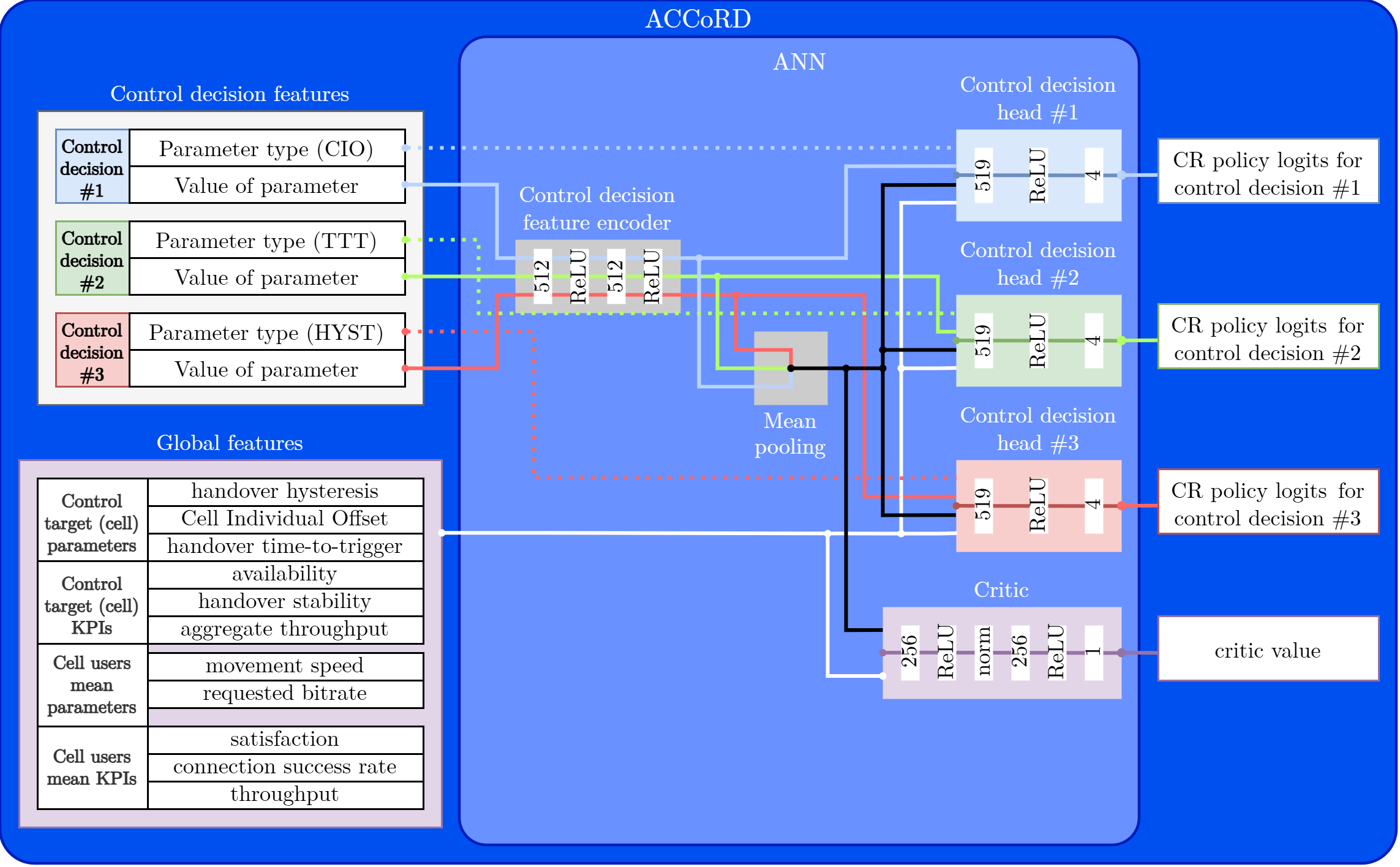}}
\caption{Structure of the actor-critic ANN implemented in ACCoRD} 
\label{fig:ann-structure}
\end{figure*}

The critic is implemented as a three-layer \ac{MLP} with \ac{RELU} activation and layer normalization applied after the first layer.
Unlike the actor heads, which focus on individual decisions, the critic evaluates the state of the entire system.
It takes the global features and the mean-pooled summary of \textit{all} encoded control decision features as input to produce the scalar state-value estimate $V(s)$.
Figure~\ref{fig:ann-structure} illustrates this structure.

\subsection{Training process}
\label{subsec:training-process}
\begin{figure}[!t]
\centerline{\includegraphics[width=0.99\columnwidth]{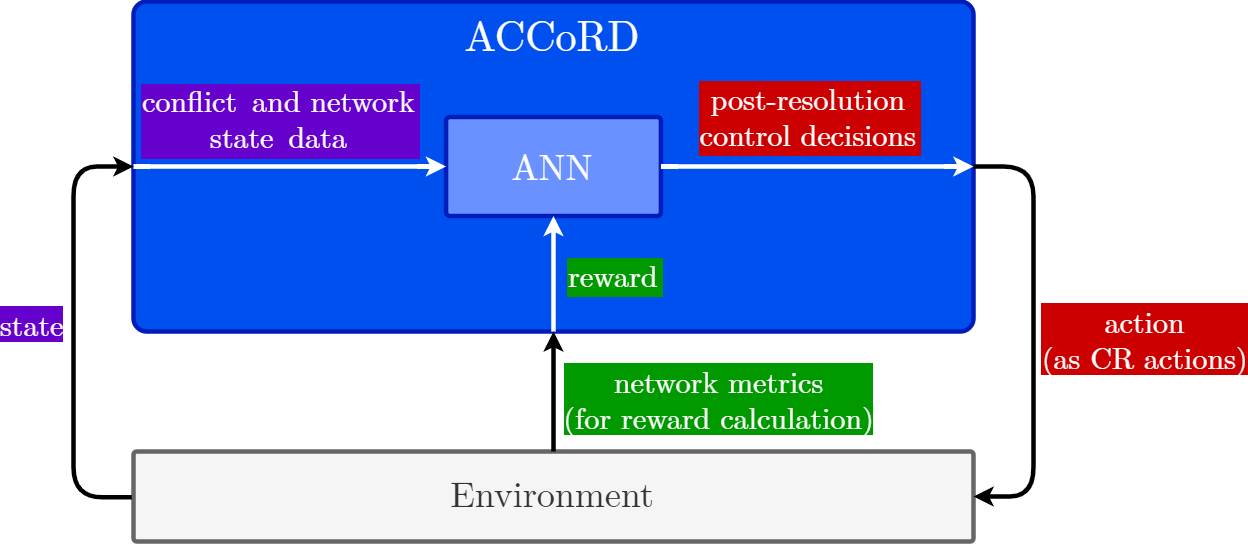}}
\caption{Data flow for ACCoRD's RL training}
\label{fig:accord-training-loop}
\end{figure}

Training utilizes the \ac{PPO}-Clip algorithm, which balances implementation simplicity with reliable convergence.
It was selected for \ac{ACCORD} because its strict bounding of policy updates provides good training stability, which is considered critical to prevent policy collapse in the highly non-stationary environment caused by the simultaneous operation of multiple uncoordinated xApps.
The "Clip" variant of \ac{PPO} stabilizes training by utilizing a clipped surrogate objective function $L^{CLIP}$~\cite{openai-ppo-2017}, defined as:
\begin{equation}
    L^{CLIP}(\theta) = \hat{\mathbb{E}}_t \left[ \min(r_t(\theta)\hat{A}_t, \text{clip}(r_t(\theta), 1-\epsilon, 1+\epsilon)\hat{A}_t) \right]
\end{equation}
where $r_t(\theta)$ is the probability ratio between new and old policies, $\hat{A}_t$ is the estimated advantage, and $\epsilon$ is the clipping hyperparameter.
The flow of state, action, and reward data during the training phase is depicted in Fig.~\ref{fig:accord-training-loop}.

Experience is gathered in $400$ s intervals. At the end of each interval, observations are split into minibatches and used to update the policy over 3 \ac{PPO} epochs. 
To ensure policy stability, \ac{KL} divergence is tracked using the approximate formula:
\begin{equation}
    D_{KL} = \hat{\mathbb{E}}_t \left[ \log \frac{\pi_{\theta_{old}}(a_t|s_t)}{\pi_{\theta}(a_t|s_t)} \right]
\end{equation}
Consecutive steps with significantly high divergence trigger learning rate adjustments or early epoch termination.
A 12-step pretraining of the critic is performed in each batch using Huber loss $L_\delta$ to improve baseline estimations:
\begin{equation}
    L_\delta(y, f(x)) = \begin{cases} 
      \frac{1}{2}(y - f(x))^2 & \text{for } |y - f(x)| \le \delta, \\
      \delta (|y - f(x)| - \frac{1}{2}\delta) & \text{otherwise.}
   \end{cases}
\end{equation}

The \ac{PPO} algorithm was configured with $3$ epochs per update, a clipping parameter of $0.10$, a target mean \ac{KL} of $0.015$, and an entropy bonus coefficient of $0.007$. Starting learning rates were set to $10^{-5}$ for the actor and $10^{-4}$ for the critic.

\subsection{Reward calculation}
\label{subsec:reward_calc}
The reward function $R_{\mathrm{network}}$ reflects the impact of \ac{CR} decisions on global network performance.
Metrics are observed in a rolling window of length $T_{\mathrm{meas}}$ at detection time ($t=0$) and post-resolution ($t=T_{\mathrm{reward}}$).
The reward is calculated as the weighted sum of changes in normalized metrics for Ping-Pong handovers ($C_{\mathrm{PP}}$), \ac{RLF} ($C_{\mathrm{RLF}}$), and Call Blockages ($C_{\mathrm{CB}}$):

\begin{equation}
\label{eq:reward}
R_{\mathrm{network}} =
-\,\frac{w^{\mathrm{rew}}_{\mathrm{PP}}\Delta \hat{C}_{\mathrm{PP}} + w^{\mathrm{rew}}_{\mathrm{RLF}}\Delta \hat{C}_{\mathrm{RLF}}
+\, w^{\mathrm{rew}}_{\mathrm{CB}}\Delta \hat{C}_{\mathrm{CB}}}
{w^{\mathrm{rew}}_{\mathrm{PP}}+w^{\mathrm{rew}}_{\mathrm{RLF}}+w^{\mathrm{rew}}_{\mathrm{CB}}}
\end{equation}

Weights ($w^{\mathrm{rew}}$) allow the operator to align the agent with specific intents. In this study, a balanced policy ($w^{\mathrm{rew}}=1.0$ for all) is used.
\section{Simulation scenario}
\label{sec:scenario}
To evaluate the efficiency of the implemented \ac{CR} Agent, a robust simulation of an \ac{ORAN}-compliant network is conducted.
The simulation scenario comprises a hexagonal arrangement of eight 5G base stations (BSes).
A single macro BS collocated with a micro BS is located in the center of the simulated area, with six evenly spaced micro BSes surrounding it.
Each micro BS is configured with three 5G cells, each covering a $120$-degree sector.
The simulation area is defined as a circle with a radius of $450$ m.


To simulate varying levels of network load, User Equipments (UEs) are distributed within the simulation area according to three distinct deployment configurations:
\begin{itemize}
    \item \textbf{Small:} 15 UEs connected to the central macro BS and 6 UEs per micro BS, totaling 57 UEs.
    \item \textbf{Medium:} 20 UEs per macro BS and 8 UEs per micro BS, totaling 76 UEs.
    \item \textbf{Large:} 30 UEs per macro BS and 10 UEs per micro BS, totaling 100 UEs.
\end{itemize}
Additionally, a single dedicated 'Reference UE' is included in each run on top of these distributions to enable granular analysis of individual connectivity parameters during simulation.
UEs exhibit diverse mobility (pedestrian and vehicular) and heterogeneous traffic profiles (low-bitrate voice, medium-bitrate data, high-bitrate data).

The simulated network hosts two xApps: \ac{MRO} and \ac{MLB}.
\ac{MRO} optimizes handover hysteresis and \ac{TTT} to minimize ping-pongs and \acp{RLF}.
\ac{MLB} controls \ac{CIO} to balance load.
Their interplay causes indirect control conflicts.

\subsection{Parametrization of ACCoRD}
The operation of \ac{ACCORD} is configured by structural and operational parameters:
\begin{itemize}
    \item \textbf{Maximum conflicting decisions ($N_{\mathrm{ConfDec}}$):} Set to $3$ (CIO, TTT, Hysteresis).
    \item \textbf{Action space size ($N_{\mathrm{A}}$):} Set to $4$.
    \item \textbf{Measurement window ($T_{\mathrm{meas}}$):} Set to $1.0$ s to align with xApp processing intervals.
    \item \textbf{Cooldown duration ($t_{\mathrm{CR}}$):} Set to $10$ s.
    \item \textbf{Training rollout ($t_{\mathrm{training}}$):} Experience is collected for $400$ s before a policy update.
\end{itemize}
\section{Methodology}
\label{sec:methodology}
The evaluation is structured around various \ac{CM} configurations to assess the impact of \ac{ACCORD}.
The baseline configuration is \textbf{\ac{CMF} inactive} (No CM).
The second configuration uses a rule-based \textbf{Prioritization with Cooldown} method, tested in two variants: prioritizing \ac{MRO} and prioritizing \ac{MLB}.
The final configuration is \ac{ACCORD}.

\subsection{Evaluation procedure}
As the \ac{ANN} is untrained upon initialization, it requires a substantial time of environment observation.
To evaluate performance, \ac{ACCORD} is first trained during 10,000 seconds of simulation.
Subsequently, a 1,000-second evaluation run is performed to assess the trained \ac{ANN}'s capabilities.
To ensure stable conditions, the initial $200$ seconds of each run are discarded.
Each configuration is tested in 36 different initial random UE placements.

\subsection{Evaluation metric}
A novel evaluation metric, the \textit{Relative Penalty Score}, is proposed to fairly rank methods across various scenarios.
First, a total penalty value $P^i$ is calculated for each simulation run by summing negative events weighted by their severity:
\begin{equation}
\label{eq:penalty_metric}
P^{i} =
w^{\mathrm{pen}}_{\mathrm{PP}}\, C^{i}_{\mathrm{PP}}
+ w^{\mathrm{pen}}_{\mathrm{RLF}}\, C^{i}_{\mathrm{RLF}}
+\, w^{\mathrm{pen}}_{\mathrm{CB}}\, C^{i}_{\mathrm{CB}}
\end{equation}
Weights are set to highlight the impact of service disruption: $w^{\mathrm{pen}}_{\mathrm{PP}}=0.05$, $w^{\mathrm{pen}}_{\mathrm{RLF}}=0.40$, and $w^{\mathrm{pen}}_{\mathrm{CB}}=0.40$.
The metric is then normalized relative to the best-performing method ($P^{\mathrm{best}}$) in the set:
\begin{equation}
\label{eq:rel_pen_metric}    
    P^{i}_{\mathrm{rel}}=\dfrac{P^{i}}{P^{\mathrm{best}}} \cdot 100\%
\end{equation}
\section{Evaluation results}
\label{sec:evaluation}
\subsection{Statistical results}
Table~\ref{tab:ran-sim-stats} summarizes the outcomes, showing the relative penalty metric averaged over 36 simulation runs.

In the \textbf{Small UE deployment}, the Prioritize \ac{MRO} method performs best ($102.5\%$), with \ac{ACCORD} close behind at $104.1\%$. The rule-based method suffices for low traffic.
However, in the \textbf{Medium UE deployment}, \ac{ACCORD} demonstrates robustness, achieving a score of $101.4\%$ and outperforming Prio. \ac{MRO} by $2.8$ pp.
The \textbf{Large UE deployment} shows even greater improvement, with \ac{ACCORD} scoring $101.6\%$ and widening the gap to the next best method to $4.0$ pp.
Prio. \ac{MLB} consistently performs the worst.
The efficiency of \ac{ACCORD} scales favorably with network load and conflict frequency, proving its value in challenging conditions.

\begin{table}[t]
\centering
\scriptsize
\setlength{\tabcolsep}{2pt}
\renewcommand{\arraystretch}{0.95}
\caption{Evaluation results based on 36 simulation runs}
\label{tab:ran-sim-stats}
\begin{tabularx}{\columnwidth}{@{} r C S[table-format=3.2] S[table-format=3.1] S[table-format=2.1] @{}}
\toprule
\textbf{Rank} & \textbf{Algorithm} &
\textbf{Avg\,Penalty} &
\textbf{Avg\,Rel.\,Penalty (\%)} &
\textbf{Delta to best (pp)} \\
\midrule
\multicolumn{5}{c}{\textbf{Small UE deployment configuration}}\\
1 & Prio.\ \ac{MRO} & 292.13 & 102.5 & 0.0 \\
2 & \textbf{ACCoRD} & 297.73 & 104.1 & 1.6 \\
3 & ConMit disabled & 302.97 & 106.3 & 3.8 \\
4 & Prio.\ \ac{MLB} & 330.43 & 116.0 & 13.5 \\
\addlinespace[2pt]
\multicolumn{5}{c}{\textbf{Medium UE deployment configuration}}\\
1 & \textbf{ACCoRD} & 402.84 & 101.4 & 0.0 \\
2 & Prio.\ \ac{MRO} & 412.82 & 104.2 & 2.8 \\
3 & ConMit disabled & 424.45 & 107.1 & 5.7 \\
4 & Prio.\ \ac{MLB} & 448.79 & 113.2 & 11.8 \\
\addlinespace[2pt]
\multicolumn{5}{c}{\textbf{Large UE deployment configuration}}\\
1 & \textbf{ACCoRD} & 549.87 & 101.6 & 0.0 \\
2 & Prio.\ \ac{MRO} & 568.56 & 105.6 & 4.0 \\
3 & ConMit disabled & 583.45 & 108.3 & 6.7 \\
4 & Prio.\ \ac{MLB} & 602.99 & 111.9 & 10.3 \\
\bottomrule
\end{tabularx}
\end{table}

\subsection{Case Study: CR policy derived by ACCoRD}
We examine a policy derived by \ac{ACCORD} in a medium deployment run where it achieved the lowest penalty.
The agent learned a strategy governed by the following rules:
\begin{itemize}
    \item \textbf{CIO stabilization:} Conflicting decisions regarding \ac{CIO} are always rejected, effectively stabilizing cell borders to prevent load oscillations.
    \item \textbf{Handover inertia:} Decisions regarding \ac{TTT} and Hysteresis are either rejected or increased by single steps. Increasing these parameters creates "inertia" in the system: higher \ac{TTT} extends the time required to trigger a handover, while higher hysteresis requires a larger signal delta. This dampens the system's reaction to transient channel fluctuations.
    \item \textbf{Overall rejection:} In nearly $64\%$ of conflicts, all conflicting control decisions were rejected.
\end{itemize}
Although the "reject-all" strategy in nearly $64\%$ of conflicts may appear conservative and could potentially block some intended xApp optimizations, it effectively acts as a noise filter that prioritizes network stability over aggressive, uncoordinated adjustments. This behavior is a direct consequence of the balanced reward policy used in this study, which heavily penalizes service disruptions. Tuning these reward weights allows MNOs to adjust the agent's risk tolerance, potentially yielding more aggressive optimization profiles. Overall, this policy minimized ping-pong handovers and call blockages (CBs) significantly compared to baselines, while keeping \ac{RLF} counts competitive, demonstrating \ac{ACCORD}'s ability to autonomously derive complex stabilization strategies.

\subsection{Computational complexity and inference latency}
\label{subsec:latency}
To ensure practical viability within the \ac{Near-RT} \ac{RIC}, where the control loop operates between $10$\,ms and $1$\,s, the \ac{CR} Agent must not introduce excessive delays. The inference time of the proposed \ac{ANN} is bounded by the $\mathcal{O}(N_{\mathrm{ConfDec}} \cdot L \cdot d^2)$ computational complexity of the feed-forward \ac{MLP} architecture, where $L$ is the number of sequential layers in the inference path and $d$ is the maximum layer width. Because the critic network is utilized exclusively during the training phase to compute advantages, it is omitted from the real-time execution path. Consequently, the runtime complexity depends solely on the shared feature encoder and the actor decision heads. Given the relatively low-dimensional state space and the lightweight architecture of \ac{ACCORD} ($N_{\mathrm{ConfDec}}=3$, $L=4$, $d=519$), a single forward pass requires approximately $3.2$ million multiply-accumulate operations. On standard CPU hardware, this executes in the microsecond to low-millisecond range. This negligible computational overhead ensures adherence to the latency requirements of the \ac{Near-RT} control loop.

\section{Summary}
\label{sec:summary}
This paper introduced \ac{ACCORD}, an \ac{ANN}-based \ac{CR} method capable of adapting its policy to network conditions.
\ac{ACCORD} utilizes a novel input/output masking scheme to handle variable conflicts and learns through \ac{PPO}-Clip to minimize a weighted penalty of negative network events.
Simulations demonstrated that while simple prioritization works for low loads, \ac{ACCORD} provides unmatched performance in medium and high traffic scenarios.

Future work will focus on:
\begin{itemize}
    \item \textbf{New control targets:} including control target type embeddings (user, slice) to handle diverse conflict scopes.
    \item \textbf{Scalability:} expanding the number of input heads to handle larger conflict sets without structural redesign.
    \item \textbf{Action space expansion:} investigating other \ac{CR} actions, including multi-step modifications or rejection without cooldown to fine-tune control.
    \item \textbf{Reward shaping and diverse KPIs:} evaluating non-uniform reward weights ($w^{\mathrm{rew}} \neq 1.0$) to reflect nuanced operator priorities, and analyzing the impact on additional metrics such as PRB utilization and spectral efficiency.
    \item \textbf{Handling rare events:} adapting the training process to handle highly imbalanced datasets, where specific conflict instances may occur in less than $10\%$ of operational time.
    \item \textbf{Real-world validation:} transitioning from simulation to hardware-in-the-loop physical testbeds to verify the solution's performance, generalization, and latency in complex, higher-dimensional over-the-air scenarios.
\end{itemize}

\bibliographystyle{IEEEtran}
\bibliography{IEEEbibliography}

\end{document}